\documentclass[runningheads]{llncs}
\usepackage[T1]{fontenc}
\usepackage{graphicx}
\usepackage{listings}
\usepackage{algorithm}
\usepackage{algcompatible}
\usepackage{algpseudocode}

\usepackage{amsmath,amssymb}
\usepackage{subcaption}
\usepackage{stmaryrd}
\usepackage{booktabs}
\usepackage{tabularx}
\usepackage{dirtytalk}
\usepackage{multirow}
\usepackage{adjustbox}
\usepackage{bm}
\usepackage{bbding}

\usepackage{makeidx}  
\usepackage[colorlinks,linkcolor=blue]{hyperref}
\usepackage{url}
\usepackage[misc]{ifsym}

\usepackage[acronym]{glossaries}
\newacronym{us}{US}{Ultrasound}
\newacronym{fps}{fps}{frames per second}
\newacronym{ef}{LVEF}{Left Ventricular Ejection Fraction}
\newacronym{es}{ES}{End-Systolic}
\newacronym{ed}{ED}{End-Diastolic}
\newacronym{sv}{SV}{Systolic Volume}
\newacronym{cdm}{CDM}{Cascaded Diffusion Model}
\newacronym{edm}{EDM}{Elucidated Diffusion Model}
\newacronym{ml}{ML}{Machine Learning}
\newacronym{mri}{MRI}{Magnetic Resonance Imaging}
\newacronym{ct}{CT}{Computed Tomography}

\newcommand{\etal}{\mbox{\emph{et al.}}}

\makeatletter
\newcommand*{\inlineequation}[2][]{%
  \begingroup
    \refstepcounter{equation}%
    \ifx\\#1\\%
    \else
      \label{#1}%
    \fi
    \relpenalty=10000 %
    \binoppenalty=10000 %
    \ensuremath{%
      #2%
    }%
    ~\@eqnnum
  \endgroup
}
\makeatother

\begin{document}
\title{Training-Free Condition Video Diffusion Models for single frame Spatial-Semantic Echocardiogram Synthesis}
\titlerunning{Training-free Echocardiogram Synthesis}
\author{Van Phi Nguyen\inst{1} \and Tri Nhan Luong Ha \inst{2} \and Huy Hieu Pham \inst{2,3} \and Quoc Long Tran \textsuperscript{$\dagger$} \inst{1} \textsuperscript{(\Letter)}} 

\institute{Institute for Artificial Intelligence, University of Engineering and Technology, Vietnam National University, Hanoi, Vietnam \email{tqlong@vnu.edu.vn} \and  VinUni-Illinois Smart Health Center, VinUniversity, Hanoi, Vietnam
\and College of Engineering \& Computer Science, VinUniversity, Hanoi, Vietnam}
\authorrunning{Phi et al.}
\def\thefootnote{$\dagger$}\footnotetext{Corresponding authors.}
\maketitle

\begin{abstract}

Conditional video diffusion models (CDM) have shown promising results for video synthesis, potentially enabling the generation of realistic echocardiograms to address the problem of data scarcity. However, current CDMs require a paired segmentation map and echocardiogram dataset. We present a new method called Free-Echo for generating realistic echocardiograms from a single end-diastolic segmentation map without additional training data. Our method is based on the 3D-Unet with Temporal Attention Layers model and is conditioned on the segmentation map using a training-free conditioning method based on SDEdit. We evaluate our model on two public echocardiogram datasets, CAMUS and EchoNet-Dynamic. We show that our model can generate plausible echocardiograms that are spatially aligned with the input segmentation map, achieving performance comparable to training-based CDMs. Our work opens up new possibilities for generating echocardiograms from a single segmentation map, which can be used for data augmentation, domain adaptation, and other applications in medical imaging. Our code is available at \url{https://github.com/gungui98/echo-free}

\end{abstract}
\section{Introduction}

Echocardiogram is a widely used imaging modality for assessing cardiac function and structure. It is a non-invasive, cost-effective, and widely available imaging modality that provides real-time information about the structure and function of the heart. Meanwhile, the interpretation of echocardiograms highly depends on the operator's experience and the quality of the images. Automated analysis of echocardiograms has the potential to improve the accuracy and efficiency of the diagnosis and treatment of cardiovascular diseases~\cite{ouyang_video-based_2020}. However, the development of machine learning models for echocardiogram analysis is challenging due to the scarcity of labeled data, the complexity of echocardiograms, and the variability of echocardiograms across different patients and imaging systems~\cite{ghorbani2020deep}.

Synthetic medical imaging therefore has the potential to address the problem of data scarcity in medical image processing. In the domain of echocardiogram, synthetic data is often generated from physics-based simulators and data-driven deep generative models. Physics-based simulators create synthetic echocardiograms by solving the wave equation, which mimics the physical process of ultrasound imaging~\cite{jensen_simulation_2004,burger_real-time_2013,garcia2022simus}. From a segmentation map, the simulator could generate an echocardiogram by simulating the ultrasound wave propagation through the tissue, typically with the help of a scattering model. However, physics-based simulators are computationally expensive and require expert knowledge to tune the parameters, such as the speed of sound, the attenuation coefficient, and the scattering coefficient, to generate realistic echocardiograms. In addition, obtaining a tissue scatter map from the segmentation map is non-trivial and often causes unrealistic textures in the generated echocardiograms. In addition to physics-based simulators, data-driven deep-generative models, such as Generative Adversarial Networks (GANs), have been proposed to generate realistic echocardiograms~\cite{salehi_patient-specific_2015,tomar_content-preserving_2021,bargsten_specklegan_2020,cronin_using_2020}. Despite promising results, the quality of the generated echocardiograms is often limited due to the mode collapse problem. Diffusion models (DMs) have recently emerged as a promising alternative to GANs for generating more realistic videos and conditions~\cite{gupta_maskvit_2022,harvey_flexible_2022,ho_imagen_2022,esser2023structure,villegas_phenaki_2022}, due to easier training procedures and better sample quality. However, controlling the generation of current DMs requires a paired dataset of segmentation maps and echocardiogram, which is not always available in the domain of medical imaging. In this work, we propose a training-free condition video diffusion model (CDM) for echocardiogram synthesis that allows the generation of realistic echocardiograms from a single end-diastolic segmentation map without the need for any additional training data.

\noindent\textbf{Contribution: } In this paper, we propose a new method for CDM for echocardiogram synthesis. Our model is training-free and requires no additional paired dataset of segmentation map and echocardiogram data, while still generating realistic echocardiograms from a single end-diastolic segmentation map. Based on SDEdit~\cite{meng2021sdedit}, we propose, Free-Echo, which start the reverse denoising process with a noisy version of a pseudo-video obtained from the segmentation map instead of pure Gaussian noise. We demonstrate the effectiveness of our model on two public echocardiogram datasets, CAMUS and EchoNet-Dynamic, showing comparable performance to training-based CDMs.

\noindent\textbf{Related work:}
\noindent\textbf{Echocardiogram synthesis: } Several methods have been proposed to generate realistic echocardiograms. Salehi \etal~\cite{salehi_patient-specific_2015} used a physic simulator to generate patient-specific echocardiograms. Liang ~\cite{liang_sketch_2022} proposed a GAN-based model to generate ultrasound from sketch image. Tomar \etal~\cite{tomar_content-preserving_2021} also used Cycle-GAN to generate content-preserving echocardiograms with unpaired data. While GAN-based methods have shown promising results for echocardiogram synthesis, the quality of generated echocardiograms is often limited due to the mode collapse problem and the difficulty of training GANs. DMs have been proposed to generate realistic echocardiograms~\cite{reynaud_feature_2023,van2023echocardiography,stojanovski_echo_2023}. Reynaud \etal~\cite{reynaud_feature_2023} developed a cascaded diffusion model, conditioned on an End Diastolic (ED) frame, to produce ultrasound images with varying left ventricle ejection fractions (LVEFs). Stojanovski \etal~\cite{stojanovski_echo_2023} utilized Denoising Diffusion Probabilistic Model (DDPM) to generate synthetic echoechocardiograms with conditions made from semantic label maps. Another DDPM-based method by Phi \etal~\cite{van2023echocardiography} applied dynamic semantic label mapping of diastolic frame in a multi-scale decoder to produce realistic echocardiography sequences with diverse anatomical structures. Diffusion-based methods have shown success, however current methods still require a paired dataset of segmentation map and echocardiogram data. In this work, we propose a training-free CDM for echocardiogram synthesis to address this problem.

\noindent\textbf{Conditional video synthesis: } CDM aims to synthesise realistic sample given some condition, from semantic description from text to spatial layout such as bouding box, segmentation map. Especially in the domain of medical imaging, CDM enables synthetic data generation with different anatomy, pathology, and acquisition parameters. With paired dataset of condition and the sample, classifier-guided \cite{dhariwal2021diffusion} and classifier-free conditioning \cite{ho_classifier-free_2022} have been proposed to the CDM on spatial layout. In some domains, such as medical imaging, where a paired dataset of conditions is not always available, researchers search for a training-free conditioning method, such as SDEdit \cite{meng2021sdedit}. However, SDEdit which requires a colorized version of the spatial layout, suffered from content leakage. In this work, we propose a training-free conditioning method based on SDEdit, that replaces the colorized version of the spatial layout with the pseudo-video obtained from the spatial layout using optimal transport.

\section{Method}

\begin{figure}
    \centering
    \vspace{-0.5cm}
    \includegraphics[width=\linewidth]{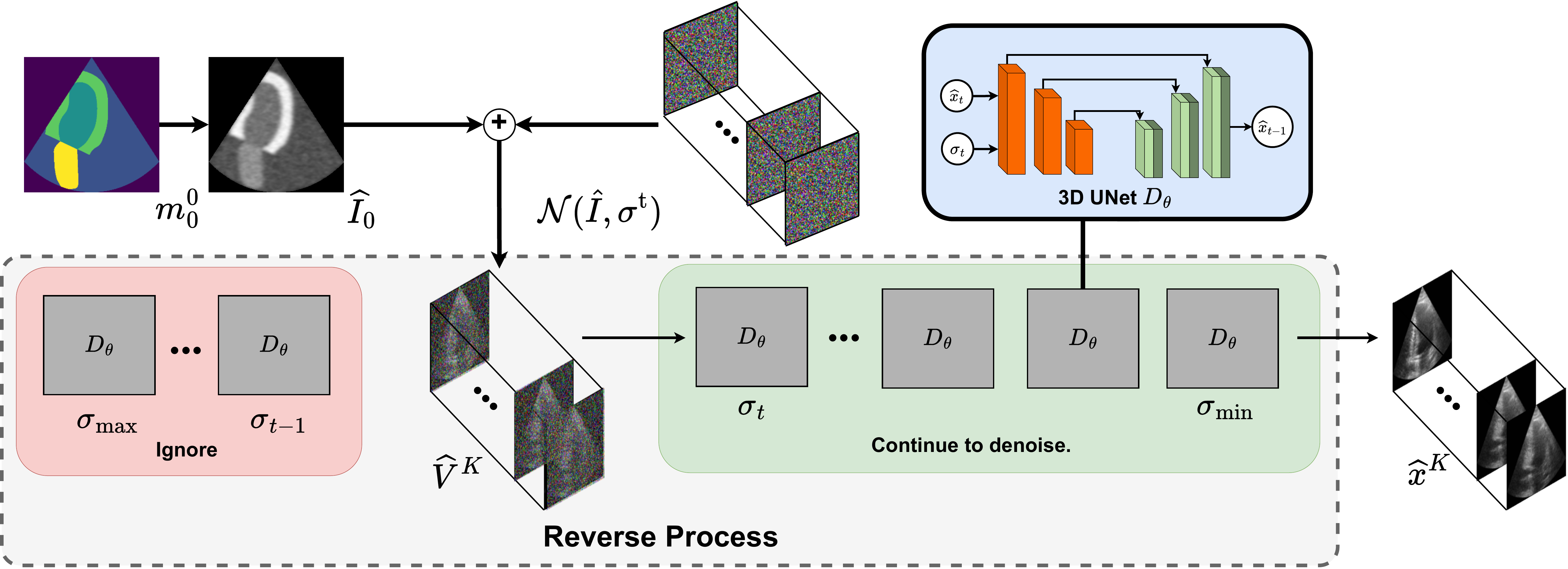}
    \caption{Illustration of our method. Given a single end-diastolic segmentation map $m_0$, we first solve the optimal transport problem to obtain the pseudo-image $\hat{I}_0$. We then start the reverse process of the DM from the diffusion step $t$ with the noisy version of the pseudo video $\hat{V}^K$, obtained by adding Gaussian noise to the pseudo image. The reverse process is continued until the diffusion step $t = 0$, at which point we obtain the generated echocardiogram $\hat{x}^K$.}
    \label{fig:narrow_model}
    \vspace{-0.5cm}
\end{figure}

\noindent\textbf{Video Diffusion Model}: DMs~\cite{sohl-dickstein_deep_2015,ho_denoising_2020} are a class of generative models that learn to transform pure Gaussian noise into data distribution by gradually removing the noise. DMs consist of two processes: the forward process and the reverse sampling process. Given the data distribution, $p(x_0^K)$ has a standard deviation of $\sigma_{\text{data}}$, where $x_0^K$ is a video volume that contains $K$ frames. The forward process creates a noisy version $s_0^K$ of the video $x_0^K$ by adding Gaussian noise of standard deviation $\sigma$ to the video volume $s_0^K = x_0^K + \sigma n$, where $n \sim \mathcal{N}(0, I)$ is a pure Gaussian noise. When the standard deviation or noise level $\sigma$ is substantially larger than $\sigma_{\text{data}}$: $\sigma_{\text{max}} \gg \sigma_{\text{data}}$ and the noisy volume $s_0^K$ is almost indistinguishable from pure Gaussian noise, the $\sigma_{\text{max}}$ is the maximum noise level.

The reverse process, used to synthesize a new video $\hat{x}_0^K$, randomly initializes the noisy video $s_0^K$ by sampling from pure Gaussian noise $\mathcal{N}(0, \sigma_{\text{max}}^2 I)$, and then gradually removes the noise from the noisy video $s_0^K$ from high noise level $\sigma_{\text{max}}$ until reach the zero noise level $\sigma = 0$, at which point we obtain the generated echocardiogram $\hat{x}_0^K$. To estimate the noise added to the noisy video $s_0^K$ at the noise level $\sigma$, the reverse process trains a denoiser $D_\theta(\cdot, \sigma)$ by minimizing the following loss function:

\begin{equation}
  \mathcal{L}(\theta) = \mathbb{E}_{\sigma \sim p(\sigma), x_0^K \sim p(x_0^K)} \left[ w(\sigma) \|D_\theta(x_0^K + \sigma n, \sigma) - x_0^K\|^2_2\right]\,,
\end{equation}
where $w(\sigma)$ is the weight function, used to balance the contribution of the different noise levels to the loss function, and $p(\sigma)$ is the noise level distribution. To sample the DMs, starting from the noisy version of the video $x$ at noise level $\sigma_{\text{max}}$, the reverse process is continued until noise level $\sigma_{\text{min}}$ by solving the following ordinary differential equation (ODE):

\begin{equation}
  \frac{d\hat{x}_t}{d\sigma} = -\sigma\Delta_{x}log(p(\hat{x}_t, \sigma)) = \frac{\hat{x}_t - D_\theta(\hat{x}_t, \sigma)}{\sigma}\,,
  \label{eq:ode}
\end{equation}
for $t \in [0, T]$ and $\sigma \in [\sigma_{\text{max}}, \sigma_{\text{min}}]$ and T is the denoising step. The solution of the ODE is obtained by the ODE solver, such as the Euler-Maruyama or 2nd order Heun's method. 

\noindent\textbf{Conditioning}: Conventionally, the denoiser $D_\theta(\cdot, \sigma)$ is modified to take condition $c$ as input, for example, the end-diastolic segmentation map $m_0^0$. The denoiser is then trained via classifier-free \cite{ho_classifier-free_2022} or classifier-guidance \cite{dhariwal2021diffusion}. In both cases, the conditioning requires a pair dataset of segmentation maps and the video. To address those challenges, we propose a training-free conditioning method based on SDEdit \cite{meng2021sdedit}. The idea is that instead of denoising pure Gaussian noise $s_0^K \sim \mathcal{N}(0, \sigma_{\text{max}}^2 I)$ to $\sigma_{\text{min}}$, at diffusion step $t_i$, we replace the noisy version of video $s_0^K$ with the noisy version of a pseudo-video obtained from the segmentation map $m_0^0$ and continue the reverse process until $\sigma_{\text{min}}$. This allows us to generate spatially coherent echocardiograms from the segmentation map while still being able to fill realistic textures with pseudo-video. In the original SDEdit, the pseudo-video is a colorized version of segmentation, which makes it harder for denoiser to map from the noisy version of the pseudo-video to the clean version of the pseudo-video. While recent research on unpaired domain-to-domain translation has shown that the pseudo-video can be generated from the segmentation map using optimal transport~\cite{su2022dual}, current works require extensive training and data from both source and target domains. In our method, we replace the colorized version of the pseudo-video with an adaptive pseudo-video generation.

\noindent\textbf{Pseudo Video Generation}: Given the segmentation map $m_0^0$, we first map each label to a unique intensity value to obtain an intensity image $\hat{I_0}$. We noticed that simple mapping may cause a mismatch in the intensity histogram between the pseudo-image and the video data. To address this issue, we solve the optimal transport problem between the pseudo-image $\hat{I_0}$ and the video data $x_0^K$ to obtain the pseudo-image $\hat{I}_0$. Formally, we solve the following optimal transport problem:

\begin{equation}
  \hat{I}_0 = \arg \min_{\hat{I}_0} \int_{\hat{I}_0} \int_{x_0^K} c(x, \hat{x}) \mu(x) \mu(\hat{x}) dx d\hat{x}\,,
\end{equation}
where $c(j, \hat{j})$ is the cost function, and $\mu(x)$ and $\mu(\hat{x})$ are the probability distribution of the video data $x_0^K$ and the pseudo video $\hat{x}_0^K$, respectively. We then use the Sinkhorn algorithm to solve the optimal transport problem with the cost function $c(x, \hat{x}) = \|x - \hat{x}\|^2_2$ and a regularization parameter of $10^{-3}$. Similar to the original SDEdit, we add Gaussian noise at step $t_i$ to the pseudo image $\hat{I}_0$ to obtain the noisy version of the pseudo video $\hat{V}^K = \hat{I}_0 + \sigma(i) n$ with $\sigma(i)$ being the noise level at diffusion step $t_i$. We then continue the reverse process of the DM in equation \ref{eq:ode} from the diffusion step $t_i$ with the noisy version of the pseudo-video $\hat{V}^K$ until we obtain the generated ultrasound video.

\noindent\textbf{Denoiser Formulation}: We follow the parametrization from  Elucidated Diffusion Model (EDM)  \cite{karras_elucidating_2022}. The noise level $\sigma$ follows the log-normal distribution $\log \sigma \sim \mathcal{N}(P_{\text{mean}}$ where $P_{\text{mean}}= - 1.2, P_{\text{std}} = 1.2)$.

The input and output of the denoiser $D_\theta(\cdot, \sigma)$ are scaled as:

\begin{equation}
\hat{D}_\theta(s, \sigma) = \frac{\sigma_\text{data}}{\sigma^\ast} s + \frac{\sigma \cdot \sigma_\text{data}}{\sigma^\ast} F_\theta\left(\frac{s}{\sigma^\ast}, \frac{\ln(\sigma)}{4}\right)\,,
\end{equation}
where $\sigma_\text{data} = 0.5$ and $\sigma^\ast = \sqrt{\sigma^2 + \sigma_\text{data}^2}$. $F_\theta$ is a neural network. The weight function $w(\sigma)$ is set to $w(\sigma) = (\sigma^\ast / (\sigma \cdot \sigma_\text{data}))^2$ that cancels the weight of $F_\theta$.

\section{Experiment Settings}
\noindent\textbf{Data: }
We evaluate our model on two datasets, CAMUS \cite{leclerc2019deep} and EchoNet-Dynamic \cite{ouyang2020video}. The CAMUS public dataset consists of 500 patient records with 2 chamber-view echocardiograms. It contains segmentation map for the left ventricle endocardium, myocardium, and left atrial endocardium. The segmentation annotation of CAMUS is available for the end-diastolic and end-systolic frames. We split the dataset into 400 videos for training, 50 videos for validation, and 50 videos for testing. The EchoNet-Dynamic dataset consists of 10,030 4 chamber view ultrasound videos. Different from CAMUS, it contains the segmentation map only for the left ventricle at the end-diastolic frame and the end-systolic frame. We split the dataset into 8,024 videos for training, 1,003 videos for validation, and 1,003 videos for testing. For both datasets, we sample the first 24 frames of the video, with condition on the end-diastolic segmentation map.

\noindent\textbf{Implementation: } We adopt the 3D-Unet from Ho et.al.~\cite{ho_imagen_2022} as our denoiser. The model is trained on 24 frames of $128 \times 128$ pixels. For the training of the denoiser, we use the Adam optimizer with a learning rate of $10^{-3}$, and a batch size of 16 for 100,000 iterations. We use the same noise level distribution as EDM \cite{karras_elucidating_2022}. We start the reverse process from the diffusion step $t_i = 15$ in a total of 64 diffusion steps. All experiments are conducted on a 3x NVIDIA H100 GPU with 80GB of memory.

\noindent\textbf{Evaluation: }
We evaluate our model on the following metrics: (1) Structural Similarity Index (SSIM) \cite{ssim}, (2) Peak Signal-to-Noise Ratio (PSNR), (3) Fr\'echet Inception Distance (FID) \cite{heusel2017gans}, (4) Fr\'echet Video Distance (FVD) \cite{unterthiner_fvd_2019} between the generated and the ground truth echocardiograms. For quantitative evaluation, we generated 10 samples for each segmentation in the test set, for FID and FVD, we use the InceptionV3 and the R(2+1)D networks, respectively. The non-deep learning metrics, such as SSIM, PSNR, are computed based on the average of the metrics between the generated and ground truth echocardiograms given the same segmentation map. We compare our model with SDEdit method \cite{meng2021sdedit} and the CDM using classifier-free conditioning \cite{ho_classifier-free_2022}, with a classifier-free guidance factor of 7.0. All models are trained on the same datasets, denoiser hyperparameters, and noise level distribution.

\noindent\textbf{Results: }

\begin{table}
  \centering
  \begin{tabular}{llc@{\hspace{10pt}}ccccc}
    \toprule
    \textbf{Dataset} & \textbf{Method} & \textbf{Step. $t$} & \textbf{SSIM $\uparrow$} & \textbf{PSNR $\uparrow$} & \textbf{FID $\downarrow$} & \textbf{FVD $\downarrow$} \\
    \midrule
     \multirow{2}{*}{CAMUS~\cite{leclerc2019deep}} &     SDEdit~\cite{meng2021sdedit}           & - & 0.23 & 19.54 & 32.45 & 213.50 \\
                            &     Cls-Free \cite{ho_classifier-free_2022}  & - & 0.52 & 22.49 & 22.50 & 150.80 \\
    \cmidrule{2-7}
                            &     Free-Echo (Ours)       & 15 & 0.48 & 20.43 & 26.31 & 195.41 \\
                            &     Free-Echo (Ours)       & 35 & 0.27 & 16.54 & 53.71 & 312.12 \\
                            &     Free-Echo (Ours)       & 55 & 0.12 & 13.45 & 67.15 & 412.55 \\
    \midrule
     \multirow{2}{*}{Echonet-Dynamic~\cite{ouyang_video-based_2020}} &     SDEdit~\cite{meng2021sdedit} & - & 0.34 & 21.52 & 25.49 & 250.45 \\
                            &     Cls-Free~\cite{ho_classifier-free_2022}  & - & 0.56 & 20.04 & 18.56 & 130.85 \\
    \cmidrule{2-7}
                            &    Free-Echo (Ours)       & 15 & 0.51 & 20.78 & 24.22 & 180.39 \\
                            &     Free-Echo (Ours)       & 35 & 0.32 & 18.72 & 44.67 & 332.12 \\
                            &     Free-Echo (Ours)       & 55 & 0.21 & 16.66 & 59.85 & 395.73 \\
    \bottomrule
  \end{tabular}
  \caption{Quantitative evaluation of our model on the CAMUS and EchoNet-Dynamic datasets.}
  
  \label{tab:quantitative}
\end{table}

\begin{figure}
    \centering
    \includegraphics[width=\linewidth]{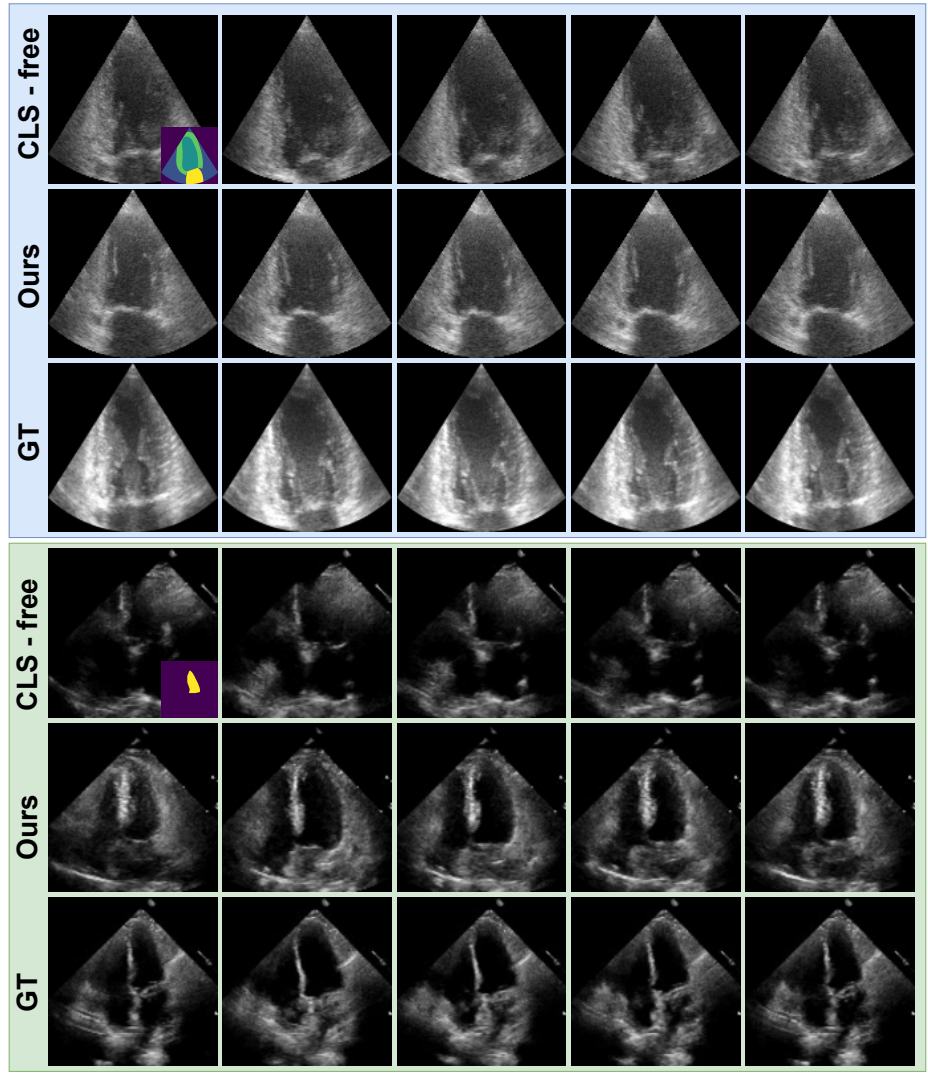}
    \caption{Visual comparison of our model with the CDM and the ground truth echocardiograms trained on the CAMUS (blue) and EchoNet-Dynamic (green) datasets.}
    \label{fig:qualitative}
\end{figure}

Table \ref{tab:quantitative} shows the quantitative evaluation of our model on the CAMUS and EchoNet-Dynamic datasets. We observe that our model achieves comparable performance to the CDM using classifier-free conditioning, with around 10\% drop in SSIM, PSNR, and L2 distance, and 20\% increase in FID and FVD. Our model outperforms SDEdit, showing the benefit of pseudo-video generation. Since there is trade-off between the diffusion step when we start the reverse process from the diffusion step $t_i$, we observe that the best performance is achieved at the diffusion step $t_i = 15$ for both datasets, and the performance decrease significantly when we start the reverse process from the diffusion step $t_i = 55$. This trade-off suggests that the high-level anatomy and motion of the heart can be generated at very early diffusion step, and the texture and fine details of the heart can be generated at later diffusion step. The performance of our model is also consistent across the two datasets. In Figure \ref{fig:qualitative}, we present a visual comparison of our model against the CDM, using ground truth data from the CAMUS and EchoNet-Dynamic datasets. For the CAMUS dataset, our model shows an ability to integrate motion that aligns with the segmentation map, even when generating from a static pseudo-video. However, the fidelity of motion diminishes in later diffusion steps, emphasizing the importance of selecting the right $t_i$ for realistic echocardiogram generation. On the EchoNet-Dynamic dataset, our model effectively synthesizes echocardiograms with realistic LV region motion, adhering to the segmentation map's anotomy. We noticed that the CDM struggles with consistent ultrasound structure, particularly in areas with artifacts and deformities like the cone area. This is likely due to the label map containing only LV region, making it difficult for the CDM to generate uniform representations. Our use of pseudo-labels in these areas helps overcome this issue, ensuring consistency in the generated ultrasound shapes. This approach shows promise in handling the inherent variability and complexity of echocardiogram data, as seen in EchoNet-Dynamic's diverse shapes.

Some limitations should be considered. The generated echocardiograms are imperfect, with low resolution and short duration. The performance of our model is also sensitive to the diffusion step $t_i$, and the optimal diffusion step may vary across different datasets. Downstream applications of our model still require further evaluation since the motion of the heart is not always consistent with the segmentation map. Future will focus on a more robust method for generating echocardiograms from a single segmentation map, and exploring the potential of our model for data augmentation, domain adaptation, and other applications in medical imaging.

\newpage

\section{Conclusion}

We propose a training-free CDM for echocardiogram synthesis. Our model can generate realistic echocardiograms from a single end-diastolic segmentation map. We demonstrate the effectiveness of our model on two public echocardiogram datasets CAMUS and EchoNet-Dynamic, and our in-house dataset. We further conduct human evaluation to assess the visual quality of the generated echocardiograms. Through extensive experiments, we show that our model can generate plausible echocardiograms with that both temporal coherent and spatially align with the input segmentation map without additional training data. Our work opens up new possibilities for generating echocardiograms from a single segmentation map, which can be used for data augmentation, domain adaptation, and other applications in medical imaging.

\begin{credits}
\subsubsection{\ackname}
The research is supported by the Vingroup Innovation Foundation (VINIF) under project code VINIF.2019.DA02. Computing resources for this research were sponsored by Intelligent Integration Co., Ltd. (INT2), Vietnam.

\end{credits}

\newpage
\bibliographystyle{splncs04}
\bibliography{Paper-1171.bib}

\end{document}


\section{Appendix}
\subsection{Models architecture details}

Table \ref{tab:unet3d_layers} shows the architecture of the Unet3D model used for all denoisers in the experiments. The model consists of 3 downsample blocks, 2 middle blocks, and 3 upsample blocks. Each block consists of a convolutional layer, followed by a spatial linear attention layer, and an attention layer for same pixel position. The number of parameters in each layer is also shown in the table. The input shape is (B, 3, K, H, W), where B is the batch size, K is the number of frames, H is the height, and W is the width. The output shape is (B, 3, K, H, W), where 3 is the number of channels (RGB). The number of channels in the input and output is 3, as the input and output are RGB images. The number of channels in the intermediate layers is 32, 64, 128, 256, and 512, respectively. The number of channels in the output of the upsample blocks is 128, 64, and 32, respectively. The final convolutional layer outputs the denoised image with the same shape as the input. The number of parameters in the model is 1,036,195. The model is implemented using the PyTorch library.

\subsection{Supplementary Videos}

In the supplementary material, we provide several videos to demonstrate the effectiveness of the proposed method. including the following:

\begin{itemize}
    \item \textbf{Compare with other methods:} We provide four videos to compare the proposed method with other methods for each dataset. The videos show the segmentation map, classifier-free condition, our method, and the ground truth. 

    \item \textbf{Effect of the denoising step:} We provide four videos to show the effect of the start denoising step for each dataset. The videos show the segmenation and result for each starting denoising step.

\end{itemize}

\begin{table}[htbp]
        \centering
        \renewcommand{\arraystretch}{1.5} 
        \caption{Description of Layers in Unet3D Model, used for all denoisers in the experiments.}
        \label{tab:unet3d_layers}
            \resizebox{\textwidth}{!}{%
                \begin{tabular}{|c|c|c|c|c|}
                        \toprule
                        Block& \textbf{Input Shape} & \textbf{Output Shape} & Layer& \textbf{Parameter Count} \\ \midrule
                        
                        \multirow{2}{*}{Initial Convolution}     & \multirow{2}{*}{(B, 3, K, H, W)}    & \multirow{2}{*}{(B, 32, K, H, W)}    & Conv3D& 672 \\ \cline{4-5}
                                                                                                         &                    &                     & Conv3D & 2112 \\ \hline 
                        \multirow{5}{*}{Downsample 1}         & \multirow{5}{*}{(B, 32, K, H, W)}   & \multirow{5}{*}{(B, 64, K, H/2, W/2)} & Conv3D & 2112 \\ \cline{4-5}
                                                                                                         &                    &                     & SpatialLinearAttention & 256 \\ \cline{4-5}
                                                                                                         &                    &                     & Attention & 576 \\ \cline{4-5}
                                                                                                         &                    &                     & Conv3D & 16512 \\ \hline 
                        \multirow{5}{*}{Downsample 2}         & \multirow{5}{*}{(B, 64, K/2, H/2, W/2)} & \multirow{5}{*}{(B, 128, K/4, H/4, W/4)} & Conv3D & 8320 \\ \cline{4-5}
                                                                                                         &                    &                     & SpatialLinearAttention & 512 \\ \cline{4-5}
                                                                                                         &                    &                     & Attention & 1152 \\ \cline{4-5}
                                                                                                         &                    &                     & Conv3D & 66048 \\ \hline 
                        \multirow{5}{*}{Downsample 3}         & \multirow{5}{*}{(B, 128, K/4, H/4, W/4)} & \multirow{5}{*}{(B, 256, K/8, H/8, W/8)} & Conv3D & 33024 \\ \cline{4-5}
                                                                                                         &                    &                     & SpatialLinearAttention & 1024 \\ \cline{4-5}
                                                                                                         &                    &                     & Attention & 2304 \\ \cline{4-5}
                                                                                                         &                    &                     & Conv3D & 264192 \\ \hline 
                        \multirow{5}{*}{Middle Block 1}       & \multirow{5}{*}{(B, 256, K/8, H/8, W/8)} & \multirow{5}{*}{(B, 256, K/8, H/8, W/8)} & Conv3D & 33024 \\ \cline{4-5}
                                                                                                         &                    &                     & SpatialLinearAttention & 1024 \\ \cline{4-5}
                                                                                                         &                    &                     & Attention & 2304 \\ \cline{4-5}
                                                                                                         &                    &                     & Conv3D & 264192 \\ \hline 
                        \multirow{5}{*}{Middle Block 2}       & \multirow{5}{*}{(B, 256, K/8, H/8, W/8)} & \multirow{5}{*}{(B, 256, K/8, H/8, W/8)} & Conv3D & 33024 \\ \cline{4-5}
                                                                                                         &                    &                     & SpatialLinearAttention & 1024 \\ \cline{4-5}
                                                                                                         &                    &                     & Attention & 2304 \\ \cline{4-5}
                                                                                                         &                    &                     & Conv3D & 264192 \\ \hline 
                        \multirow{5}{*}{Upsample 1}           & \multirow{5}{*}{(B, 512, K/4, H/4, W/4)} & \multirow{5}{*}{(B, 128, K/4, H/4, W/4)} & Conv3D & 8400 \\ \cline{4-5}
                                                                                                         &                    &                     & SpatialLinearAttention & 512 \\ \cline{4-5}
                                                                                                         &                    &                     & Attention & 1152 \\ \cline{4-5}
                                                                                                         &                    &                     & Conv3DTranspose & 264192 \\ \hline 
                        \multirow{5}{*}{Upsample 2}           & \multirow{5}{*}{(B, 256, K/2, H/2, W/2)} & \multirow{5}{*}{(B, 64, K/2, H/2, W/2)} & Conv3D & 2112 \\ \cline{4-5}
                                                                                                         &                    &                     & SpatialLinearAttention & 256 \\ \cline{4-5}
                                                                                                         &                    &                     & Attention & 576 \\ \cline{4-5}
                                                                                                         &                    &                     & Conv3DTranspose & 66048 \\ \hline 
                        \multirow{5}{*}{Upsample 3}           & \multirow{5}{*}{(B, 128, K, H, W)}     & \multirow{5}{*}{(B, 32, K, H, W)}     & Conv3D & 672 \\ \cline{4-5}
                                                                                                         &                    &                     & SpatialLinearAttention & 64 \\ \cline{4-5}
                                                                                                         &                    &                     & Attention & 144 \\ \cline{4-5}
                                                                                                         &                    &                     & Conv3DTranspose & 16512 \\ \hline 
                        \multirow{2}{*}{Final Conv}           & \multirow{2}{*}{(B, 64, K, H, W)}     & \multirow{2}{*}{(B, 3, K, H, W)}      & Conv3D & 195 \\ \cline{4-5}
                                                                                                         &                    &                     & Conv3D & 672 \\ \bottomrule
                \end{tabular}%
                }
\end{table}